\documentstyle[preprint,aps]{revtex}
\tightenlines
\begin{document}
%%\draft
\title{Unconventional metallic state\\
in two dimensional system with broken inversion symmetry}
\author{V.\ M.\ Pudalov}
\address{Institute for High Pressure Physics, Troitsk, \\
Moscow district 142092, Russia}
%\date{1 June 1997}
\maketitle

\begin{abstract}
We present a model that explains two phenomena,
recently observed in high-mobility Si-MOS structures: (1) the strong
enhancement of metallic conduction at low temperatures, $T<2$\ K, and
(2) the occurrence of the metal-insulator transition in 2D electron
system. Both effects are prescribed to the spin-orbit interaction
anomalously enhanced by the broken inversion symmetry of the confining
potential well.

\end{abstract}

\pacs{PACS numbers: 71.30.+h, 73.40.Qv, and 73.20.Fz}

{\bf Introduction.}
Recently, in experiments with high mobility Si-MOS structures,
a strong drop in resistivity $\rho(T)$
has been found \cite{krav94} as temperature decreases below
$\simeq 2$\ K. This effect is evidently in disagreement with
the conventional interpretation of the one-parameter scaling theory (OPST)
\cite{abra79}, according to which all states in 2D system at
zero magnetic field should be localized  in the limit of $T\rightarrow 0$.
The subsequent scaling analysis of the temperature
and electric field dependencies of the conductivity \cite{krav95}
has revealed a critical behavior, typical for the
metal-insulator transition.
Finally, convincing evidence for the existence of the extended states in
2D system at zero field has been obtained in experiments
{\em in magnetic field}, in studies of the quantum Hall
effect to insulator transitions  \cite{puda94}.
The extended states which in high magnetic field are located in the
centre of the corresponding Landau bands, at decreasing field were found
to remain in a finite
energy range, giving rise to a mobility edge.

The experimental results thus suggest
the existence of a true metallic state and of
the metal-insulator (M-I) transition
in 2 dimensions. These results are in apparent
contradiction with the conventional OPST, and
the origin of the metallic state remains puzzling.
In this work, both experimental findings
are explained as a consequence of the spin-orbit
interaction enhanced by
the broken inversion symmetry.
The suggested model provides a good agreement
with the experimental data on the temperature
dependence of the resistivity $\rho(T)$.

{\bf Analysis of the experimental results.}
Fig.\ 1 shows a set of the curves $\rho$ vs $T$
typical for the high-mobility samples \cite{krav94},
at different electron densities $n_s$. At $T\gtrsim 2$\ K,
the resistivity, $\rho$, increases slowly as temperature decreases,
the latter is characteristic for the weakly localized regime.
At lower temperature,  $\rho$ drops sharply
for all curves belonging to the
``metallic'' range of densities, $n_s > n_c$.

The resistance drop is observed at densities in the range
from $n_c$ to $\approx 1.4\times n_c$.
The critical density, $n_c$ is sample dependent
and is equal to
$9.2\cdot 10^{10}$\ cm$^{-2}$ for the sample shown in Fig.\ 1.
The drop in $\rho (T)$ diminishes
with decreasing sample mobility,
and is almost replaced by a
conventional rise in $\rho$ at $T\rightarrow 0$
in the sample with 8 times lower mobility,
$\mu=5000$ cm$^2$/Vs.
The latter behavior is consistent with that reported in earlier
studies on low-mobility samples \cite{pepper}.

{\bf Empirical fit of the data.}
The $\rho (T)$- curves in Fig.\ 1 may be fitted well by
an empirical dependence which summarizes
the scattering probabilities of two processes:

\begin{equation}
\rho(T) = \rho_0 + \rho_1 exp(- T^*/T).
\end{equation}

The first term is independent of temperature, while the second one
describes a scattering through an energy gap, $\Delta = k_B T^*$.
The curves shown in Fig.\ 1 by thick continuous lines were
obtained using two fitting parameters for each density,
$ \rho_1 /\rho_0$ and  $T^*$.

{\bf Possible microscopic mechanisms.}
As seen from Fig.\ 1, the characteristic
temperature, $T^*$, is of the order of $2$ K
in high mobility samples.
Searching for a proper microscopic mechanism, we find two small
energy gaps intrinsic to Si-MOS structures at $H=0$: ~the
valley splitting  $\Delta _v\approx 2.4$ K and the
zero-field spin-gap $\Delta _s(H=0) \approx 3.6$ K \cite{puda85}.

It seems attractive to link the resistance drop to the
transitions between the two electron valleys,
located close to the X-points in the Brillouin zone
\cite{ridley}.
However, the intervalley "um-klapp"
scattering would hardly occur, since it requires
a combination of the reciprocal lattice vectors of the very high order.
On the other hand, the phonon-induced
intervalley transition would require participation of high
energy phonons, $E_{ph}\sim 10^4$K, and is
therefore unlikely at low temperatures.
Electron tunneling, as the intervalley transition mechanism,
would not lead to a strong temperature
dependence of scattering.

{\bf Spin-orbit splitting and interaction effects.}
In the one-electron approximation, the spin-orbit interaction is described by
the Hamiltonian \cite{ridley}:

\begin{equation}
H_{so} = \frac{\hbar^2}{4m^2c^2}[\nabla V({\bf r}) {\bf \times p}] {\bf \sigma},
\end{equation}

where ${\bf p}$ and ${\bf \sigma}$  are the momentum and spin operators,
correspondingly. For the 2D electron system in Si, the contribution of
the bulk crystal potential in $\nabla V$ is small ($g^* \approx 2$),
and the lack of inversion symmetry of the triangular confining
potential $U(z)$ plays the major role.

This lifts  the spin-degeneracy at zero
magnetic field, and leads to the appearance of a linear term
in the energy spectrum of 2D electrons \cite{rashba}:

\begin{equation}
E^{\pm }(k)=\frac{\hbar ^2k^2}{2m^{*}} \pm \alpha k,
\end{equation}
The corresponding spin-gap

\begin{equation}
\Delta_s = E^{+} - E^{-} = 2\alpha k_{F},
\end{equation}

can be viewed as the difference in energy for the electron states with
spin directed in the plane but to the left and right side with respect
to ${\bf k}_F$, or, equivalently, along and opposite
to the effective magnetic field
${\bf H^*} \sim (1/m^{*}c)\left[ {\bf k_F\times} \nabla U \right] $
in the frame related to electrons moving in 2D plane with Fermi velocity.

We suggest that the  empirically determined energy gap, $\Delta$
originates from $\Delta_s$, and is equal to  $ p\, \Delta_s$
(where $p \sim 1$), whereas the temperature independent
contribution to the resistivity, $\rho_0$, is related
to the spin-independent scattering.
Finite quantum relaxation time $\tau _q$~and the corresponding
level broadening $\hbar /\tau _q$~ should reduce the effective
gap:

\begin{equation}
k_B T^* =  p\, \Delta_s - \hbar /\tau _q.
\end{equation}

{\bf Comparison of the model with the experimental results.}
The spin splitting $\Delta _s(H=0) = 3.6$~K
for the same Si-MOS structures was determined
from the magnitude of the quantum oscillations of the
chemical potential  \cite{puda85},  extrapolated to zero field.
It is clear from Eqs. (3) and (4), that this value corresponds to
$E^{-}_{min} = - 2m^{*} \alpha/ \hbar ^2$.
With this result we obtain $\alpha = 1.79 \cdot 10^{-5}$\,K\,cm \,
and the total energy spectrum becomes known.

In order to estimate the quantum level broadening, $\Gamma =\hbar /\tau _q$,
the temperature dependence of the diagonal resistance was measured
in the quantum Hall effect regime with Fermi energy adjusted to
the Zeeman energy gap at $\nu = 6$
in the field of $H=0.75$ T. As a result, we have obtained an estimate,
$\Gamma = 2.8$ K for $n_s = 10.84\cdot 10^{10}$cm$^{-2}$
(here we presumed the Gaussian level broadening and
$\Gamma$ to be the full width).
Comparing the  model effective energy gap,
$\Delta = p\, \Delta_s - \Gamma$
with the empirical value, $T^* =0.96$\, K  obtained in the fit
at $n_s = 10.84\times 10^{10}$  (the 5th curve from the bottom in Fig.\ 1),
we eventually find $p = 0.46$ \cite{remark2}.

The empirical energy gap, $k_B T^*$,
decreases to zero at $n_s = n_c$, as seen in Fig.\ 1.
In the above model, Eq. (5), this occurs  because (i)
the level broadening increases, $\Gamma \propto 1/n_s$
and (ii) the spin splitting,
$\Delta_s = 2 \alpha  k_F$, diminishes  $\propto n_s^{1/2}$.
The empirical fitting parameter $\Gamma$  shown in Fig.\ 2
vs electron density, rises indeed as density decreases.
The total fit in Fig.\ 1
is in surprisingly good agreement with the experimental data,
despite the very simplified character of the above model.

As $n_s$ decreases and approaches $n_c$,
the resistance drop starts at lower temperatures.
The weak decrease in  $\rho (T)$
noticeable at $T>4$ K is presumably due to the weak
localization corrections
$\delta \rho \propto -\log (T/T_0)$ \cite{alts85}.
This effect was ignored in the above
model and the $\rho (T)$ points corresponding
to the negative $\partial \rho/\partial T$ were not fitted;
these points are connected by dashed lines in Fig.\ 1.

Fig.\ 2 shows also two relaxation times:
 $\tau _p$  calculated from the mobility in the  $T=0$ limit,
and $\tau _q=\hbar /\Gamma $. It is noteworthy that  both
 $\tau _p $ and $\tau _q$  diminish almost linearly as density decreases
(but at $n_s > n_c$)
and independently of each other. The momentum relaxation
time $\tau _p$  provides the necessary resistivity value,
$\rho \sim h/e^2$ at the critical density \cite{krav94,krav95},
whereas  $\tau _q$  provides the effective spin-gap equal to 0
at $n_s=n_c$.
As a result,  $\tau _p$ {\it decays  faster and becomes smaller
than} $\tau _q$  with decreasing $n_s$.
At the critical density, $\mu $ is of the order of 0.1 m$^2$/Vs and
$\Gamma \simeq 5$\,K, that corresponds to
$\tau_p =0.17\times 10^{-12}$\,s, and
$\tau _q \simeq 1.5\times 10^{-12}$\,s.
The conclusion on interception of the two curves,
$\tau_q (n_s)$ and $\tau_p (n_s)$ is model independent,
whereas the numerical values
of $\tau _q (n_s)$ and the interception point
depend on the  model chosen for level broadening.
For instance,  the interception occurs at 11.5 or
$13\cdot 10^{10}$ cm$^{-2}$ for Gaussian or Lorenzian
broadening correspondingly.

{\bf Metal-Insulator transition at $H = 0$:
Spin-orbit interaction and symmetry effects.}
It is not only the low-temperature resistance drop
but the total scaling behavior
strongly depends on the symmetry. The corresponding
universality classes of the symmetry  for random systems
were established by Dyson \cite{dyson}.
In the presence of the spin-orbit (SO) interaction, the orthogonal
symmetry of the system is replaced by the symplectic symmetry.
Correspondingly, the level-repulsion exponent in the random matrix
statistics \cite{dyson} changes from $\omega =1$ to $\omega =4$.
It appears therefore, that states are less easily localized
in systems with large $\omega$.

The effect of the spin-orbit interaction on  weak localization
was studied both theoretically and
experimentally \cite{pikus96a}.
For the strong localization regime, there have been
suggestions \cite{ando89} that M-I transition can occur
in the presence of a strong SO interaction.
The scaling function in 2D was found
to behaves asymptotically
like $\beta (G) \sim -a/G$ in the high conductance limit
$G\gg 1$, with $a>0$ in the orthogonal
and $a<0$ in the symplectic case \cite{hikami}. The $\beta $\
-function in the symplectic case may thus become
positive at sufficiently large $G $.
As disorder increases and conduction $G$ decreases,
all states will be localized even in the symplectic case. The
critical level of disorder and the critical conduction $G_c$
correspond to the point at which $\beta (G_c)=0$.

The behavior of the symplectic $\beta(G) -$ function in 2D
is qualitatively consistent with the experimental data presented
in Fig.\ 1, in the vicinity of $n_c$.
As temperature, or broadening increases,
the energy relaxation time $\tau_{\epsilon}$
appears as a cut-off parameter and $1/\tau _{\epsilon} $ may become larger
than the inverse spin relaxation time
$1/\tau _\epsilon \gg 1/\tau _s$. Then the system would again
behave as in the orthogonal symmetry case.
The natural measure of the SO interaction strength is the spin-orbit
gap $\Delta _s$ given by Eqs. (4) and (5), whereas as an estimate for
disorder we adopted $\hbar /\tau _s \approx \hbar /\tau _q = \Gamma $.
Thus, one may expect the M-I transition would manifest in those samples
where~ $\Delta _s/\Gamma \geq 1$, which is also consistent with occurrence
of the transition in the samples with peak mobility larger than
5000 cm$^2$/Vs \cite{puda96}.

{\bf Discussion.}
The above model explains why
the resistance drop is seen only in low-disordered samples
with large $\tau _q$-values \cite{puda96}.
The effect is dependent also on the symmetry of the potential well.
This provides a key for testing the driving mechanism.
As for other systems, the zero field spin-gap in GaAs/Al(Ga)As
is smaller by a factor of $10\div 100$, due to the smaller
$g^{*}$-factor value and much smaller $\nabla U$ \cite{eise84}.
Thus, even in ideal samples with zero
broadening, the resistance drop may occur at temperatures
$10\div 100\times $
lower than those for the Si-MOS structures.
In accord with this, no signatures of the resistance
drop were revealed in recent measurements on GaAs/Al(Ga)As
heterojunctions at temperatures down to 20mK   \cite{dahm96}.

Recently, there have been suggestions on other possible
collective mechanisms, such as Coulomb interaction
\cite{piku96}, spin-triplet pairing \cite{beli97},
and non-Fermi-liquid behavior \cite{dobr97}. However,
the corresponding models
are not developed yet to provide a comparison with the
experimental data.

{\bf Summary.}
It seems likely that the recently observed metal-insulator transition
in high-mobility Si-MOS structures is the first experimental manifestation
of the spin-orbit interaction induced transition in 2D.
The enhancement of the metallic conduction in
these samples at low temperatures  fits  the same framework.
Strong SO interaction energy relative to the level broadening,
and broken inversion symmetry are favorable for
the 2D metallic state. The Coulomb interaction in this model
provides the small level broadening at density down to $n_c$.
In the recent experiments \cite{simo97,puda97},
the 2D metallic phase was found to
be easily destroyed by the in-plane magnetic field;
this is a strong evidence for the spin-related origin of the 2D metal.

\section{Acknowledgments}
Author acknowledges hospitality of the Institut f\"{u}r Halbleiter Physik at
the Universit\"{a}t Linz, where the work was partly performed.
Author appreciates valuable discussions with G. Bauer, G. Brunthaler,
M.V. Entin, V. Kravtsov, I.M. Suslov,  E.I. Rashba,
and V. Volkov.
The work was supported by
the Russian Foundation for Basic Research (grant 97-02-17387),
by the Programs on "Physics of solid-state nanostructures" and
"Statistical physics" and by grant from NWO the Netherlands.

\begin{figure}[tbp]
\caption{Typical temperature dependencies of the resistivity for
a high-mobility sample \protect\cite{krav94,krav95}.
Electron density, from bottom to the top, is
equal to  13.69,  12.81,  12.15,  11.50,  10.84,  10.18,  9.53
8.87, 8.21, 7.55, 7.12 $\times 10^{10}$ cm$^{-2}$
\protect\cite{krav94,krav95}.
 Solid lines - simulations, as discussed in the text.}
\end{figure}

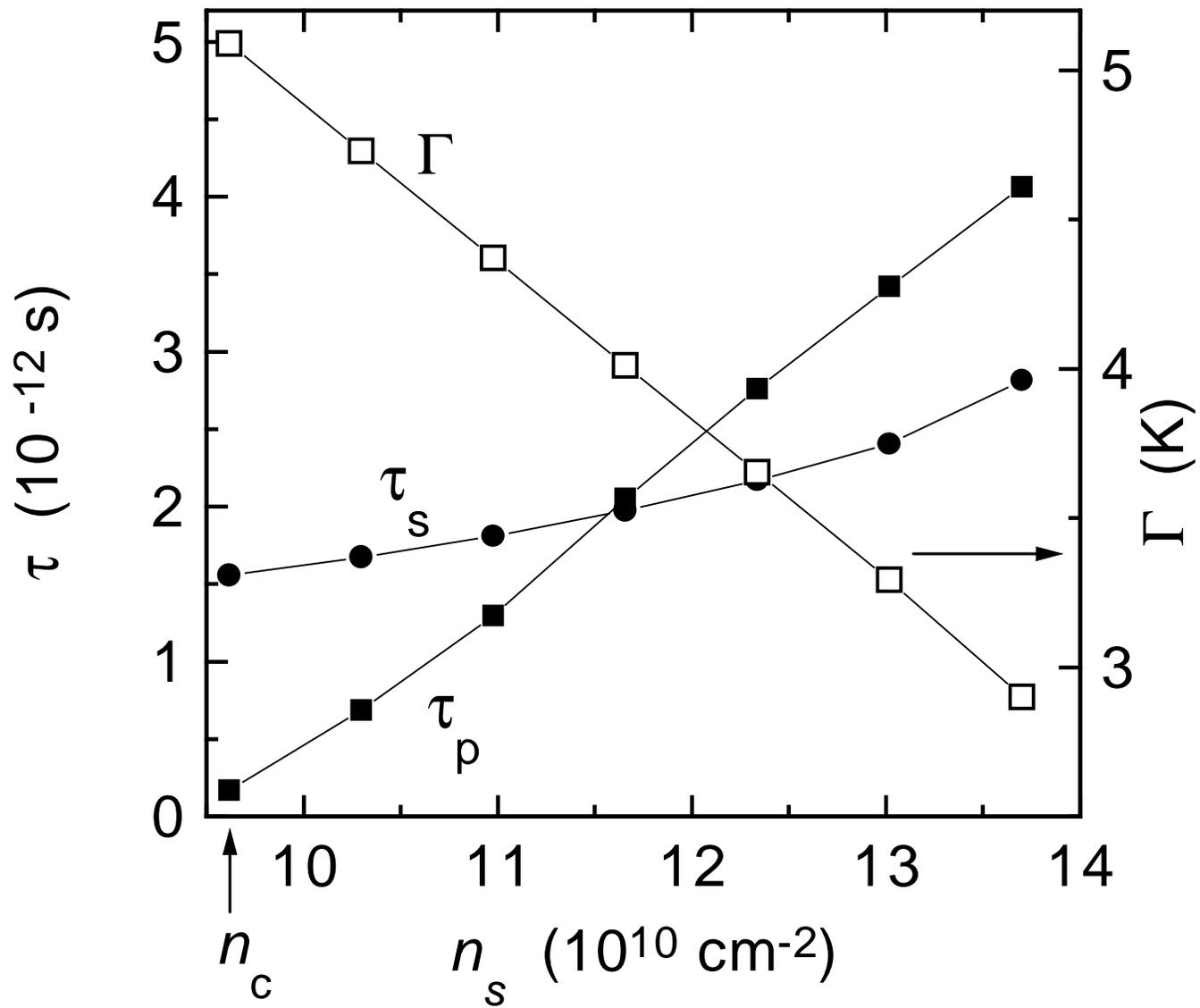
\begin{figure}[tbp]
\caption{Level broadening, $\Gamma$, and relaxation times,
as determined from the fit, Fig.\ 1,  in the vicinity of $n_c$.}
\end{figure}


\begin{references}

\bibitem{krav94}  S.V. Kravchenko, G.V. Kravchenko, J.E.
Furneaux, V.M. Pudalov, and M. D'Iorio, Phys. Rev. B {\bf 50}, 8039
(1994).

\bibitem{abra79}  E. Abrahams, P.W. Anderson, D.C. Licciardello, and
T.V. Ramakrishnan, Phys. Rev. Lett. {\bf 42}, 673 (1979).

\bibitem{krav95} S.V. Kravchenko, W.E. Mason, G.E. Bowker, J.E.
Furneaux, V.M. Pudalov, and M.D'Iorio, Phys. Rev. B {\bf 51}, 7038
(1995). \, S.V. Kravchenko, D. Simonian, M.P. Sarachik,
W. Mason, and J.E. Furneaux, Phys. Rev. Lett. {\bf 77}, 4938 (1996).

\bibitem{puda94} V.M. Pudalov, M.D'Iorio, J.W. Campbell, Surf.
Science {\bf 305}, 107 (1994).

\bibitem{pepper}  M.J. Uren, R.A. Davies, M. Kaveh, and M. Pepper,
J. Phys. C: Solid State Phys. {\bf 14}, 5737 (1981).

\bibitem{puda85}  V.M. Pudalov, S.G. Semenchinskii, and V.S. Edel'man,
ZhETF {\bf 89}, 1870 (1985); [Sov.Phys. JETP {\bf 62}, 1079 (1985)].

\bibitem{ridley}  B.K. Ridley, {\em Quantum Processes
in Semiconductors}, Clarendon Press, Oxford, 1993.

\bibitem{rashba}  Yu.A. Bychkov and E.I. Rashba, Pis'ma ZhETF,
{\bf 39}, 66 (1984); [JETP Lett. {\bf 39}, 78 (1984)].

\bibitem{remark2} From the classical viewpoint, the $p=0.5$ value
means that the average scattering angle is $\approx \pi /2$ for transitions
between  $\pm$ branches of the spectrum (2).

\bibitem{alts85}  B.L. Altshuler, A.G. Aronov, in: {\em Electron Electron
Interaction in Disordered Systems}, A.L.Efros and M.Pollak
eds., North-Holland, Amsterdam, 1985.

\bibitem{dyson}  F.J. Dyson, J. Math. Phys., {\bf 3}, 140 (1962);
{\bf 3}, 157 (1962); {\bf 3}, 166 (1962); {\bf 3}, 1191 (1962);
{\bf 3}, 1199 (1962).

\bibitem{pikus96a} F.G. Pikus, G.E. Pikus, Cond-mat/9606108. \,
W. Knap, G.E. Pikus, F.G. Pikus et al. Phys. Rev B {\bf 53}, 3912 (1996).

\bibitem{ando89} T.Ando, Phys. Rev B {\bf 40}, 5325 (1989).

\bibitem{hikami} S. Hikami, A.I. Larkin, Y. Nagaoka, Progr.\,Theor.\,Phys.,
{\bf 63}, 707 (1980). \, S. Hikami, {\em ibid} {\bf 64}, 1425 (1980).

\bibitem{puda96} V.M. Pudalov, p.34 in: Proc. Intern. Conf. on
Electron Localization and Quantum Transport in Solids, Jaszowiec,
Poland, 1996. Ed. by T.\,Dietl, Inst. of Physics PAN, Warsaw (1996).

\bibitem{eise84} J.P. Eisenstein, H.L. St\"{o}rmer, V. Narayanamurti,
A.C. Gossard, and W. Wiegmann, Phys. Rev. Lett. {\bf 53}, 2579 (1984).

\bibitem{dahm96}  F.W. Van Keuls, X.L. Hu, H. Mathur, H.W.
Jiang, and A.J. Dahm, {\it ibid}, p.33.

\bibitem{piku96}  F.G. Pikus, A.L. Efros, {\it ibid}, p.10.

\bibitem{beli97} D. Belitz, T.R. Kirkpatrick, Cond-mat/9705025.

\bibitem{dobr97} V. Dobrosavljevic, E. Abrahams, E. Miranda,
and S. Chakravarty, Cond-mat/9704091.

\bibitem{simo97} D. Simonian, S.V. Kravchenko, and M.P. Sarachik,
Cond-mat/9704071.

\bibitem{puda97} V.M. Pudalov, G. Brunthaler, A. Prinz, and G. Bauer,
Pis'ma ZhETF, {\bf 65}, 887 (1997).


\end{references}
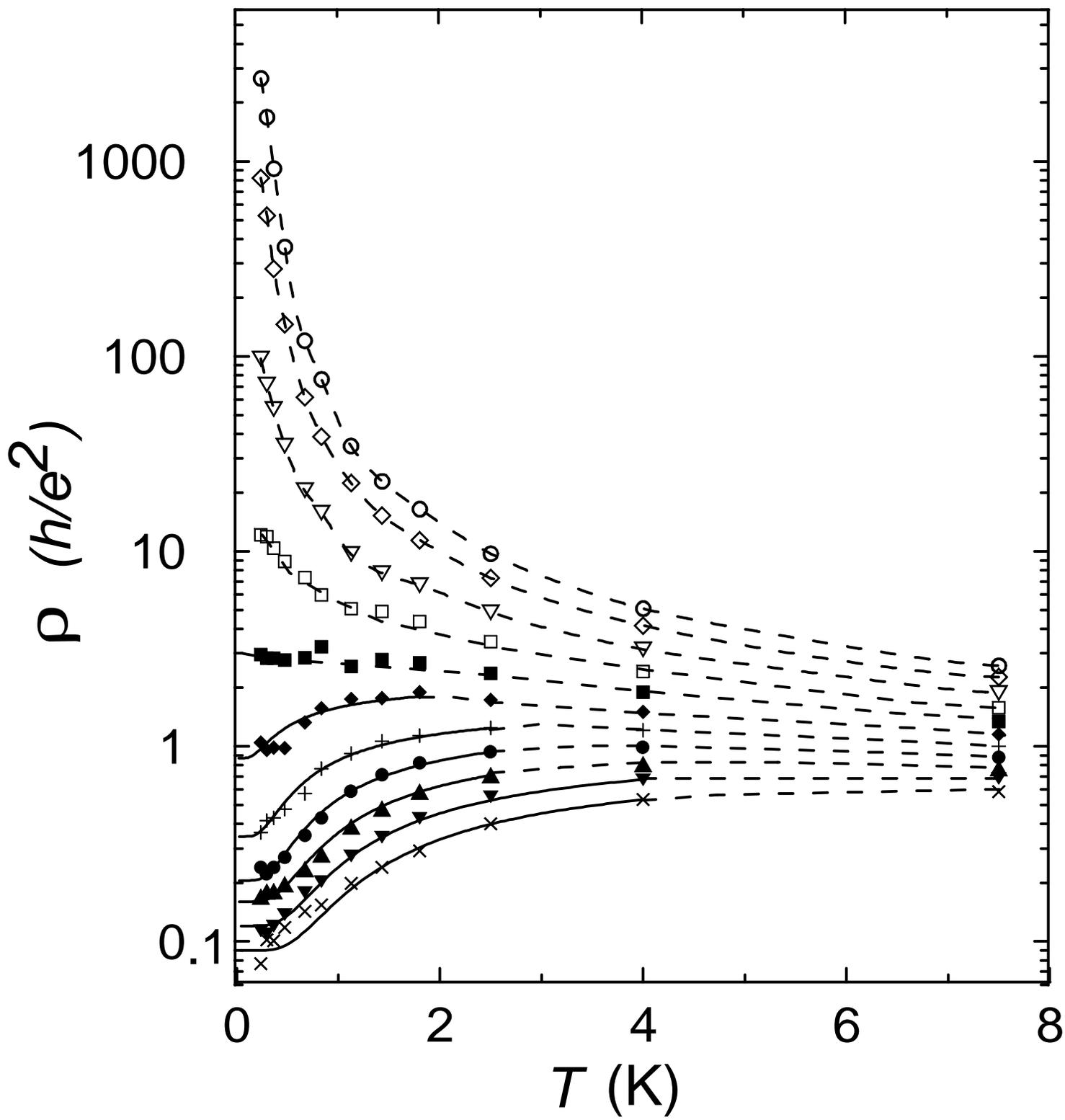
\end{document}